\documentclass{icrc2009}

\usepackage{graphicx}   
\usepackage{fixltx2e}
\usepackage{url}

\newcommand{\shorttitle}[1]%
{\markboth{Proceedings of the 31\MakeLowercase{$^{st}$} ICRC, {\L}\'{o}d\'{z} 2009}{#1} }

\begin{document}
\title{Status of the EAS studies of cosmic rays with energy below $10^{16}$eV}

\author{\IEEEauthorblockN{Hongbo Hu\IEEEauthorrefmark{1}}
                            \\
\IEEEauthorblockA{\IEEEauthorrefmark{1}Key Laboratory of Particle Astrophysics,
 Institute of High Energy Physics,CAS, Beijing 100049, China}}

\shorttitle{Hongbo Hu Cosmic Rays at Knee}
\maketitle

\begin{abstract}

This paper briefly summarizes the status of the cosmic ray observations by EAS (Extended Air Shower)
experiments with energy below $10^{16}$eV and the related studies of the hadronic interaction
models. Based on the observed sharp knee structure and the irregularities
of the cosmic ray spectrum around knee energy, plus the newly discovered electron and positron
excess, the origin of the galactic cosmic rays and the single source model  interpretation are discussed,
but  convincing evidence is not yet  available. High precision measurements of the mass composition of
primary cosmic rays at knee energy will be very useful to disentangle the problem. To reach
this goal, a better understanding of the hadronic interaction models is crucial.
It is good to see that more dedicated accelerator and cosmic ray experiments will be conducted soon.
As one EAS component, the muon distribution and muon charge ratio are important for
testing the hadronic interaction models. In addition muons are an important background to
neutrino experiments and all underground ultra-low background experiments. They are also a very useful
tool for the meteorological studies.
\end{abstract}


\section{Introduction}
This paper is a writeup of the rapporteur talk at the 31st International Cosmic Ray Conference,
 July 15,2009 in Lodz/Poland.  The talk and this paper cover the works submitted to the HE session
 for cosmic ray energy below $10^{16}$ eV. The contributions include 26 papers in HE1.1 (observation and simulation
 at energies less than $10^{15}$ eV), 16 papers in HE1.2
 (observation and simulation at energies of about $10^{15}$ - $10^{16}$ eV),
 15 from HE1.5 (muons in EAS), 16 from HE1.6 (new experiments and instrumentation)
 and 20 in HE2.1 (particle interactions relevant for cosmic ray studies with energies less than $10^{16}$ eV).

 Despite the large success of the observations of high energy $\gamma $ rays during the last twenty years,
 direct evidence for the acceleration of VHE cosmic ray nuclei has not yet been convincingly obtained.
 Therefore, it has become consensus that from the point of view of gamma ray observations we must go to PeV energies
 ~\cite{Aharonian} where the hadronic process should become visible.
 New activities following this approach are underway recently
 (e.g.,~\cite{icrc0297},~\cite{icrc0869}).
 Strictly speaking, gamma ray observations can only
 directly be related to
 the origin and acceleration mechanisms
 of the newly generated cosmic rays,
 observation on the cosmic ray particle themselves is the most fundamental and
 probably the only way to resolve the questions related to the acceleration of cosmic rays at ``early'' time.
 These cosmic rays compose the vast majority of the galactic cosmic rays as we can see them today. Many new
 observations, new understanding, new ideas and plans for the future were presented in this conference;
 it is impossible for me to cover all works and all details neither in the rapporteur talk
 nor in this paper. I do apology for the incompleteness of this summary paper.
 
\section{Observations and interpretations of cosmic rays around knee energy}

Cosmic rays were discovered by V.Hess in 1912 and have contributed continuously and greatly
 to our knowledge of particle physics and astrophysics. However, the origin, acceleration
 and propagation of cosmic rays remains unresolved since nearly hundred years (~\cite{Hillas}, ~\cite{Butt}).
 The cosmic ray spectrum exhibits almost a power law structure from $10^9$ eV to the highest energy of
 $10^{20}$ eV with an index of about -3. The spectrum has a few subtle features; the first
 one occurs at about 4PeV where the spectrum changes from -2.7 to -3.1. This break of spectrum was
 discovered in 1958 ~\cite{knee} and is known as the ``knee''.

 It is generally believed that cosmic rays with energy below $10^{16}$ eV are accelerated
 at the shock wave in SNRs, but possibly also by other astrophysics environments in the
 galaxy(~\cite{icrc0306} and refs. therein).
 The knee has been regarded in this picture as an indication of the maximum energy to
 which cosmic rays can be accelerated by the shock wave in SNRs.
  Besides, many other alternative explanations do exist.
 The knee may come from the relatively faster leakage of PeV cosmic rays into the extragalactic
 space compared to lower energy cosmic rays; or it may be due to threshold interactions, such as e+e- pair production
 by PeV cosmic rays interacting with the infrared radiation photon background at source
 (~\cite{IHEP_e_knee} and refs. therein);
 or yet some exotic interactions in the EAS development,
 where undetectable particles are produced when cosmic rays are
 above PeV energy (~\cite{horandel} and refs. therein).

\begin{figure*}[!t]
 \centering
 \includegraphics[width=13cm,angle=0]{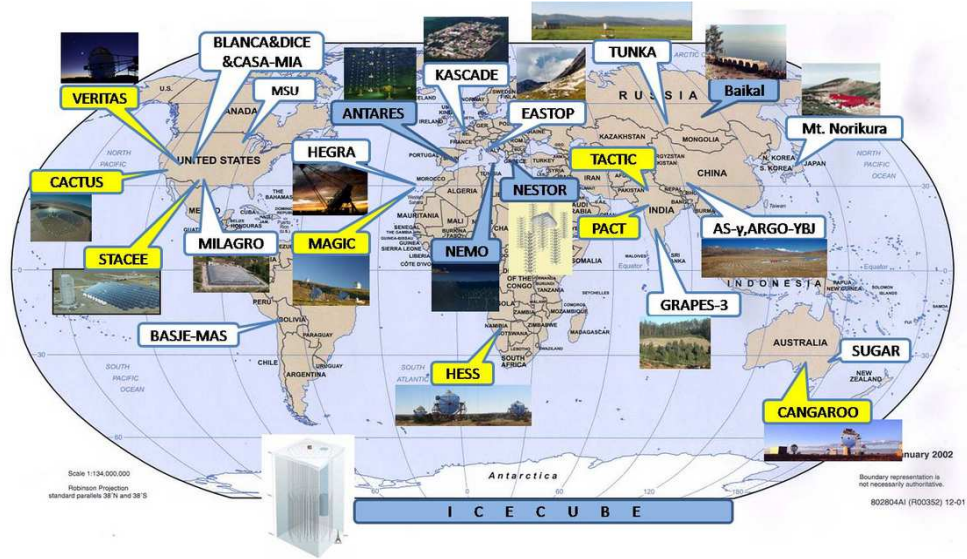}
 \caption{\label{fig1EASDet}World-wide distribution of EAS experiments
          with sensitive energy around knee energy.
          Meaning of colors: White -
          traditional EAS experiments, yellow - air Cerenkov
          telescope experiments, blue - deep
          water or deep ice experiments. }
 \label{simp_fig}
\end{figure*}
\begin{figure}[!t]
  \centering
  \includegraphics[width=8cm,height=6cm,angle=0]{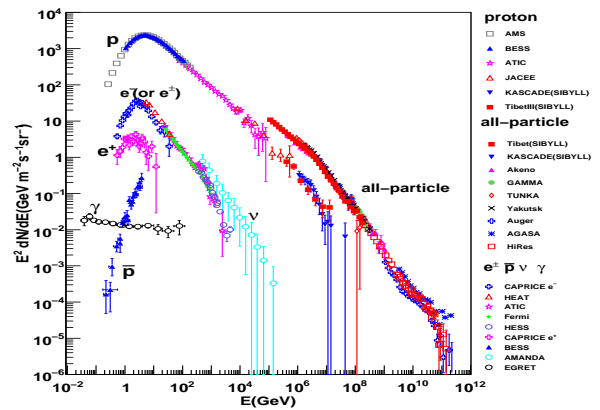}
  \caption{\label{fig2Spectra} Compilation of the cosmic ray spectra, updated with respect to the compilation
  by T.K.Gaisser in {~\cite{gaisser2006}} . Cited data: 
proton, AMS~\cite{AMS}, BESS~\cite{BESS98}, ATIC~\cite{ATIC-1}\cite{ATIC-2}, JACEE~\cite{JACEE}, KASCADE(SIBYLL)~\cite{KASKADE(SYBYLL/QGSJET)}, TibetIII(SIBYLL)~\cite{ASG_proton}; 
all-particle, Tibet(SIBYLL)~\cite{ASG_knee}, KASCADE(SIBYLL)~\cite{KASCADE}, Akeno~\cite{Akeno}, GAMMA~\cite{GAMMA}, TUNKA~\cite{TUNKA}, Yakutsk~\cite{Yakutsk}, Auger~\cite{Auger}, AGASA~\cite{AGASA}, HiRes~\cite{HiRes};
$e^{\pm}\bar{p}\nu\gamma$, CAPRICE e-~\cite{CAPRICE_electron}, HEAT~\cite{HEAT}, Fermi~\cite{Fermi}, HESS~\cite{HESS2008,HESS2009}, CAPRICE e+~\cite{CAPRICE_electron}, BESS~\cite{BESS}, AMANDA~\cite{AMANDA}, EGRET~\cite{EGRET}.} 
 \end{figure}
\begin{figure}[!ht]
  \centering
  \includegraphics[width=9cm,height=6cm,angle=0]{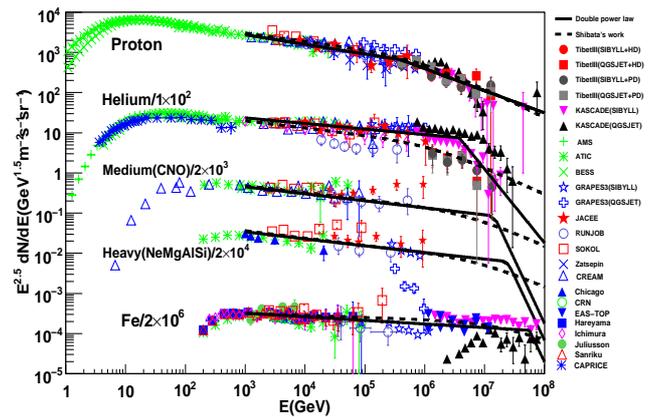}
  \caption{\label{fig3}Compilation of cosmic ray spectra for different mass groups. The Solid lines are obtained
   assuming a double power law spectrum for each mass group. For the case of proton,the RUNJOB and Tibet AS$\gamma$
   results are the input to the fit, and for the Helium, the JACEE and KASCADE (SIBYLL) are the input to the fit.
   As for all the other nuclei, parameters in (~\cite{horandel}) are used to describe the spectra before break,
   and all nuclei simply share one power law index parameter after the break ($\gamma_{2nd}$).
   The break position is assumed to be proportional to the mass of the nuclei, such as $A \times E_{cut}$.
   Both   $\gamma_{2nd}$ and $E_{cut}$ are fitted to best describe the all particle spectrum above the knee.
   The dashed line are fitted from individual nuclei as described in (~\cite{icrc0294},~\cite{icrc0295}). 
   Cited data: TibetIII(SIBYLL+HD),TibetIII(QGSJET+HD),TibetIII(SIBYLL+PD) and TibetIII(QGSJET+PD)~\cite{ASG_proton}, KASCADE(SIBYLL) and KASCADE(QGSJET)~\cite{KASCADE},
AMS~\cite{AMS}, ATIC~\cite{ATIC-1,ATIC-2}, BESS~\cite{BESS98}, GRAPES3(SIBYLL) and GRAPES3(QGSJET)~\cite{icrc1461}, JACEE~\cite{JACEE}, RUNJOB~\cite{RUNJOB}, SOKOL~\cite{SOKOL}, Zatsepin~\cite{Zatsepin}, CREAM~\cite{CREAM}, Chicago~\cite{Chicago}, EAS-TOP~\cite{EAS-TOP}, Hareyama~\cite{Hareyama}, Ichimura~\cite{Ichimura}, Juliusson~\cite{Juliusson}, Sanriku~\cite{Sanriku}, CAPRICE~\cite{CAPRICE}. } 
  \label{simp_fig}
 \end{figure}

 Because of the low event rate at knee energies, the spectrum measurements were performed so far only by ground
 based experiments and with data taken over many years of operation.
 Fig.\ref{fig1EASDet} is a brief compilation of the world wide distribution of ground based EAS experiments
 in operation in the past or at present
 with sensitive energy covering the knee region.
 Some of these experiments achieved remarkable progress in recent years.
 Firstly the energy resolution was improved and the statistics increased for the all-particle spectrum
 measurement around the knee energy (~\cite{icrc0301} and refs. therein).
 Second, several experiments extended the range of their energy spectrum and therefore make it
 possible to connect their own spectrum with the one measured from balloon-borne experiments at low energy
 and from UHECR experiments at the highest energy end, which in turn assures that
 the absolute energy scale is properly calibrated for each individual experiment.
 Thirdly, at lower energy the electron spectrum is measured and new features are uncovered at about 1 TeV energy
  (~\cite{ATIC},~\cite{PAMELA},~\cite{HESS2008},~\cite{HESS2009},~\cite{Fermi}),
 which has an impact on high energy cosmic ray physics.
 As a brief summary, Fig.\ref{fig2Spectra} shows a compilation of currently available spectra for
 all particles and for proton, antiproton, electron,positron, gamma ray and atmospheric neutrino.
 Spectra for other nuclei are shown in Fig.\ref{fig3}.
 
 Many good review articles are available in the literature on the subject of cosmic rays observation.
 One most recent example can be found in (~\cite{bluemer2009} and refs. therein).
 In this conference,
 new measurements of the spectrum and composition of cosmic rays around the knee energy are presented for
 various detector techniques (~\cite{icrc0407}, ~\cite{icrc0970}, ~\cite{icrc0712}).
 In general, all-particle spectra of those experiments agree
 reasonably well with previously published results. The situation for composition measurements has not
 much improved and remains quite model dependent
(~\cite{icrc1461},~\cite{icrc0214}, ~\cite{icrc0421},~\cite{icrc0573},~\cite{icrc0737},~\cite{icrc1213},~\cite{icrc1429}).
 As an example, using the muon multiplicity
 information, GRAPES-3 studied the primary composition up to PeV energies with SIBYLL and QGSJET models.
 Significant differences are seen between the two model assumptions.
 They found that SIBYLL model better describes data than
 QSGJET; the former one can make the individual spectra better agree with the direct measurement.
 The same conclusion was obtained when comparing the composition results
 of the Tibet AS$\gamma$ with the KASCADE experiment ~\cite{ASG_proton}.
 Following the successful measurement of the Fe spectrum from 13-200TeV with HESS telescope ~\cite{HESS_Fe},
 VERITAS reported their preliminary Fe spectrum measurement with the same direct Cerenkov light detection
 method and demonstrated that the new approach is rather promising for composition study up to PeV energies
 ~\cite{icrc0671}.

 As for the multi-TeV cosmic rays  anisotropy, what had been observed by the Tibet AS$\gamma$
 experiment ~\cite{ASG_anisotropy}
 is now confirmed by the ARGO-YBJ experiment ~\cite{icrc0814}. The
 stability of the anisotropy is reported by the Tibet AS$\gamma$ experiment with data taken over 9 years which
 almost covers half of a solar cycle ~\cite{icrc0827}. The geomagnetic effect on the cosmic rays
 anisotropy has been studied for cosmic rays from 50 TeV to 5 PeV ~\cite{icrc0574}.
 Though the effect is at
 10\% level, it can be well removed from the anisotropy measurement as this is a stable effect and only
 relevant for the horizontal coordinate system.

 Preliminary results on the antiproton/proton ratio at few TeV energies is reported by
 ARGO-YBJ ~\cite{icrc0432} and a sensitivity to exclude some of the current antiproton
 direct-production models is expected soon.

 The Tibet AS$\gamma$  has recently measured the all-particle spectrum from
 100TeV to 100PeV with improved statistics and analysis ~\cite{ASG_knee}.
 Located at 4300 m above sea level, an altitude where the shower development almost
 reach its maximum for cosmic rays around the knee energy, the Tibet AS$\gamma$ experiment can detect more
 secondary particles and thus has a very good resolution for the shower size (about 5\% at PeV energy).
 As a matter of fact, the dominant contribution to the energy resolution (about 17\% at PeV energy) comes from the
 interaction models and composition models, which demonstrates
 that the systematic uncertainty becomes the
 main source of experimental error for spectrum measurements around the knee energy.
 Nevertheless, with 55M events collected from  November 2000 to October 2004, the
 experiment was able to observe a sharp knee structure with the world highest significance ~\cite{icrc0301}.

\begin{figure}[!ht]
  \centering
  \includegraphics[width=8cm,height=6cm,angle=0]{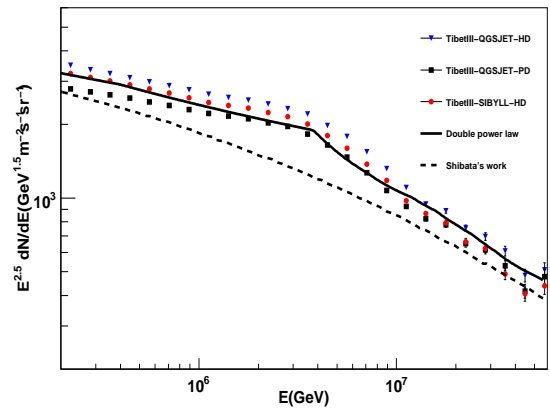}
  \caption{\label{fig4} Example of the sharp knee structure in all-particle spectrum as measured by 
   the Tibet AS$\gamma$ with different model assumptions ~\cite{ASG_knee}.
   Solid and dashed lines are the summation of the fitted spactra shown in Fig.\ref{fig3}, respectively.}
 \end{figure}

 By simply considering that individual cosmic ray spectra follow a power law with a
 rigidity-dependent exponential cut off, the spectrum of individual nuclei can be
 fitted with a set of data measured from both direct and indirect experiments
 (represented by the dashed curves in Fig.\ref{fig3} ). Though
 the summation of the fitted individual spectra may well agree with the measured all-particle spectrum
 before and after the knee, the sharp knee structure measured by the Tibet AS$\gamma$ experiment
 can not be reproduced (as shown by the dash curve in Fig.\ref{fig4} ) and another component is
 therefore required.
 It is found that such an additional component should have an index very close
 to -2 before the knee and with a very fast cut-off after the knee. Such a hard spectrum is surprisingly 
 consistent with the expected cosmic rays spectrum at the source and indicates the possible existence 
 of a nearby cosmic rays source(~\cite{icrc0294},~\cite{icrc0295}).

 \begin{figure*}[!ht]
  \centering
  \includegraphics[width=8cm,height=15cm,angle=-90]{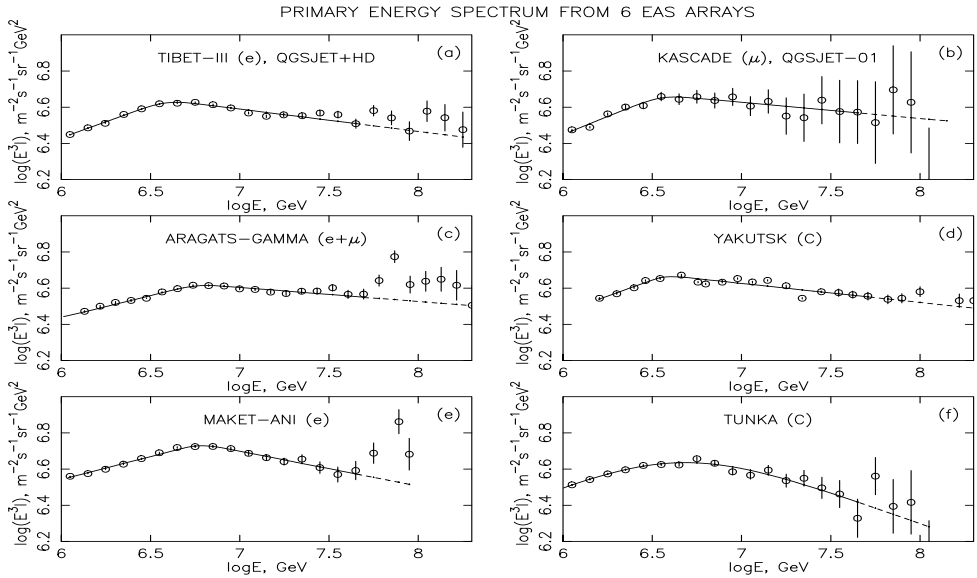}
  \caption{\label{fig5} All-particle spectrum covering the knee energy range as
                         measured recently by 6 experiments
                     (a:Tibet-III,b:KASCADE, c:GAMMA, d:Yakutsk, e:Maket-Ani and f:Tunka ).
                          The line shows the
                         fitted spectrum with a free sharpness parameter as described in
                         ~\cite{icrc0301}. First 5 spectra exhibts sharp knee structure,
                         and all 6 measured spectrum show deviation from the fitted one at high energy.
                         The plot is taken from ~\cite{icrc0301} where more details can be found.}
 \end{figure*}
 One might argue that the above mentioned choice of the smooth exponential cutoff spectra
 for individual nuclei and a lower Helium spectrum, which
 is in favour of
 the results obtained by RUNJOB and Tibet AS$\gamma$, has lead to the necessity of the additional
 component. By adopting the sharp break double-power-law spectra and allowing the Helium to be the main
 component at the knee energy according to the measurement made by JACEE ~\cite{JACEE},
 ATIC(~\cite{ATIC-1,ATIC-2}),
 CREAM ~\cite{CREAM}
 and KASCADE (SIBYLL) ~\cite{KASCADE}, the sharp
 knee spectrum can be well reproduced, as shown by the solid lines in Fig.\ref{fig3}-\ref{fig4}.
 Comparing the solid and dashed line in Fig.\ref{fig3}, we see that the main difference between
 the two fits is the different treatment of the experimental Helium spectra. And
 this demonstrates again the complicated situation in the composition measurement and the consequence
 for the physics interpretation.

\begin{figure}[!t]
  \centering
  \includegraphics[width=7cm,height=8cm,angle=-90]{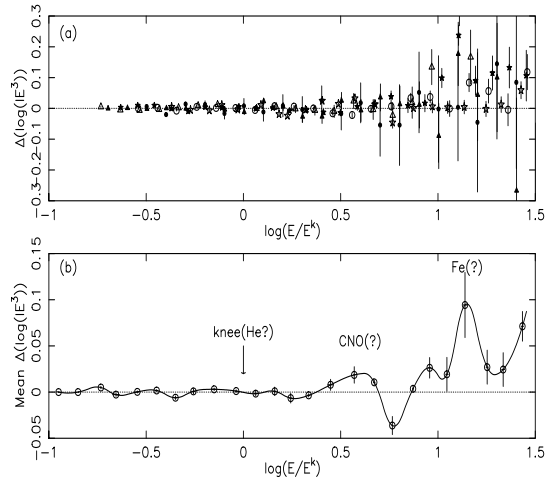}
  \caption{\label{fig6} The deviation of the measured spectrum from the fitted one
                        (a) before and (b) after the combination of all 6 experiments.
                        See ~\cite{icrc0301} for details.}
  \label{simp_fig}
 \end{figure}

 The sharp knee structure has been observed by many more experiments (Fig.\ref{fig5}) and naively favors the single nearby
 source explanation(~\cite{icrc0301} and the refs. therein).
 However the observation of the unexpected small cosmic ray anisotropy above 100 TeV
 requires a fine tuning of the source direction and cosmic ray intensity in order to make it cancel
 with the effect from the diffusion of the cosmic rays in the galaxy ~\cite{EW_anisotropy}.
  Nevertheless, if single nearby
 source is the right explanation, the sharp knee should not be the only feature in the spectrum around
 the knee and a careful analysis of the spectrum around the knee would be necessary. By redefining a
 dimensionless energy in respect with the knee position, the difference between the absolute energy
 scale of individual experiments can be removed, and the deviation of the observed spectrum from the fitted
 spectrum can be combined for all experimental data.
 With improved statistics, the combined result exhibits
 rather clearly the peculiarities at the position expected for CNO group and Fe group if the knee
 corresponds to the position of Helium (Fig.\ref{fig6}). From 
this point of view, the irregularities also support the
 single nearby source model ~\cite{icrc0301}.

 As we mentioned earlier,
 the recent electron and positron observations made by ATIC, PAMELA, HESS and Fermi
 show clearly the
 electron and positron excesses from 10GeV-1TeV, a cut off spectrum at 1 TeV is also evident. This
 feature of the electron spectrum agrees reasonably well with the single source prediction made a few
 year ago ~\cite{EW_e_cutoff}.
 This agreement is in turn regarded as a new evidence in support of the single source
 model. On the other hand,the excess and cut-off of the electron and positron spectrum reminds
 that similar behavior appears in the cosmic rays spectrum whenever a threshold interaction is 
involved, such as in the case of pion production leading to GZK spectrum (~\cite{GZK_1},~\cite{GZK_2})
 and the e+e- pair production
 leading to the dip (and bump) spectrum at EeV energies ~\cite{EeV_dip}.
 One recent work ~\cite{IHEP_e_knee}
 suggests to explain simultaneously
the electron/positron excess and the knee of the cosmic ray spectrum by introducing the pair production
 of electrons and positrons via the interaction between PeV cosmic rays and ambient infrared radiation
 background photons at the source location. Interestingly speaking, such a picture can also naturally
 explain the sharp knee structure owing to the pileup of events at the break points, which bump up
 the spectrum and make the knee apparently sharper after summing over contributions from all nuclei.
 One might guess that these pileups may be able to account as well for the irregularities mentioned in
 last paragraph, but a more quantitative calculation is needed.

 In short, no matter what the true reason behind those phenomena is, we feel it safe to conclude that
 either there exists one dominant source, maybe nearby or maybe not so nearby but strong (e.g., the
 galactic center can be one such candidate as has been proposed in ~\cite{GC1},~\cite{GC2},~\cite{GC3} ),
 or there
 exists one highly standard source, which dominates the production of the galactic cosmic rays. To
 clarify this issue, we need to better measure the composition of cosmic rays around the knee energy.
 To reach this goal, we should first improve our knowledge on hadronic interactions.

\section{Test of hadronic interaction models }
 The importance of the hadronic interaction models can well be understood through the schematic diagram
 presented in ~\cite{icrc0884}. As has been explained in the previous section, spectrum and composition
 of cosmic rays around the knee energy have to be measured by the ground based experiments, which measure
 only a profile of the extended air shower at observation level. The experimental data are the
 convolution of the primary cosmic ray distributions with all the  effects related to the atmospheric
 cascading, which is governed by the hadronic interaction and electromagnetic interaction. When doing
 the data analysis, one needs to de-convolute the data and extract the primary cosmic ray information.
 Thanks to the quantum electrodynamic theory, electromagnetic cascading can be well described, but
 unfortunately non-perturbative calculations for the QCD theory is in practice not possible, and hence the calculation
 of the hadronic interaction in EAS has to rely on phenomenological models. Currently, a few hadronic
 interaction models are available and their agreement with observations keep improving.
 In general, those models extrapolate - for the purpose of cosmic ray research
 - the results obtained by accelerator experiments for the central rapidity region to the forward
 rapidity region. The validation of this extrapolation needs to be tested and further improved.

 Whereas the LHCf ~\cite{icrc0212} and TOTEM experiments will study the very forward region physics
 with proton-proton beams at very high center of mass energy
 of about 100PeV, which is directly relevant to the longitudinal shower profile, NA61-SHINE
 ~\cite{icrc1095} studies the forward region at much low energies which are important to the lateral
 distribution of shower particles at ground level. NA61-SHINE had a successful run in 2007 with
 proton-carbon collisions at 31GeV and is scheduled to take data in 2009 with 31GeV proton beam as well
 as a pion beam at 158 and 350 GeV on carbon target. With a large acceptance and good particle identification 
capability in the forward region, those measurements will cover the cosmic ray phase space from knee energy
 up to ultra-high energies, from the point of view of the last hadronic interactions
happening in the shower. Low energy hadronic interaction models are also tested ~\cite{icrc1321}
with proton and antiproton fluxes measured by BESS  experiment and Mt.Norikura experiment.
While FLUKA can describe better the proton spectrum than other two models, UrQMD and GHEISHA
on the contrary, show much better agreement with antiproton spectrum than FLUKA. In conclusion, more experimental
results are necessary to clarify the confused situation.

\begin{figure}[!t]
  \centering
  \includegraphics[width=8cm,height=7cm,angle=0]{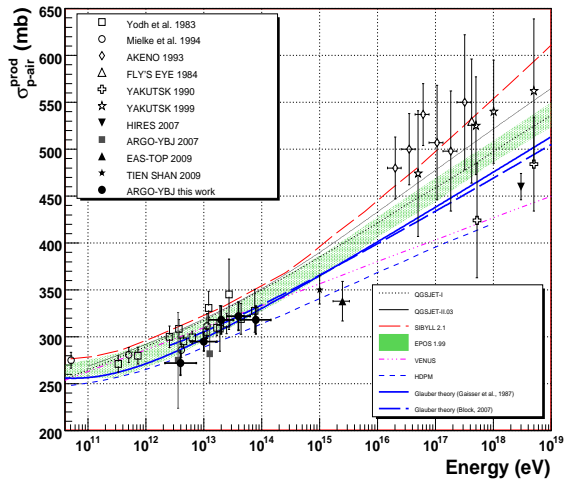}
  \caption{\label{fig7}Proton-air interaction cross section measured by various cosmic ray experiments
           together with the values given by several hadronic interaction models.
           Also shown are the predictions of two
           different calculations based on Glauber theory applied to
           accelerator data (see ~\cite{argo_xsec} and refs. therein). }
 \end{figure}

As a complementary, cosmic rays experiments are also contributing to improving the hadronic
 interaction models. New measurements of the cross section between cosmic rays and atmospheric nuclei
 are presented by the ARGO-YBJ ~\cite{icrc0319} and TienShan experiments ~\cite{icrc0222}.
 Both experiments measured a smaller cross section of proton-Air interaction
 than what had been adopted in models.
 The new results are well consistent with the
 conclusions obtained by the model developers. By comparing the model calculation and KASCADE
 observation, both QGSJET ~\cite{icrc0227}
and EPOS ~\cite{icrc0428} prefer a lower cross section between proton and air.
 Fig.\ref{fig7} shows a compilation of the proton-air interaction cross section measurements and their
comparison with calculations and values used in models. 

Forward region interactions could also be studied by cosmic ray experiments if there would not be the
 so called ``entanglement'' problem, which stands for the embarrassing situation that EAS experiments
 have only shower information at ground level measured but the two convolution gradients, i.e., the 
primary composition and hadronic interaction, need to be extracted. As presented in (~\cite{icrc0483},~\cite{icrc0875}) ,
 the authors
 propose to resolve the problem by testing and refining the model at lower energy of about tens of TeV
 where the composition of cosmic rays has been  unambiguously measured by balloon experiments. After 
which, one can study the knee composition at higher energy with the improved models. Another interesting
 proposal ~\cite{icrc0066}
 is to build a hybrid experiment, which combines a Cerenkov telescope with a traditional EAS
 array (e.g., to move KASCADE-Grande to HESS site).
 The telescope can be used to detect the direct Cerenkov light emitted from heavy nuclei
 as the light yield is proportional to the charge square of the cosmic ray particle. With this signal,
 a high purity Fe sample can be selected, and with the information recorded by the EAS array the hadronic
 interaction models can be studied. Given the fact that no plan has been proposed to study the
 forward region physics by heavy nuclei collisions in accelerator experiment, this idea seems to be
 very important to cosmic ray research at very high to extremely high energy.

Electromagnetic cascading has been much better understood except the LPM effect, which predicts a smaller
 cross section at high energy and in dense medium due to the multiple scattering. LPM effect is not only
 important to the high energy electron-positron observation, but is also important for high energy hadronic
 interactions and neutrino physics. Experimentally, the LPM effect was clearly demonstrated ~\cite{icrc0836}
 for electron
 energies above about 50GeV. Those new results might be helpful in  understanding the discrepancy between
 the ATIC and Fermi electron/positron spectrum.

Being the first and largest carpet structure detector in the world, and owing to its very fine granularity
 in temporal and spacial measurement, the ARGO-YBJ experiment presented preliminary results on the 
time structure of the extensive air shower front ~\cite{icrc0424}
 distribution and multi-core event search ~\cite{icrc0682}.
 In general, data agree with model prediction, but more careful
 work will be necessary.

So far, we have assumed that no new physics is necessary to explain the knee structure or other unusual
 events observed in cosmic rays around the knee energy. In case that new physics is involved
 in the formation of the knee spectrum  (e.g., new heavy particles, resonance states, quark gluon plasma etc),
 it is suggested in ~\cite{icrc0884} that 
 the muon energy measurement will be unavoidable in order to understand the full
 story of the knee. The same paper says that a Russian-Italian complex NEVOD-DECOR, which uses the
 water Cerenkov technique, is about ready to do this measurement, yet a standard EAS array is preferred
 to be added to the complex.

\section{Muon physics}
Muons are one important EAS component, which carries rich information about the properties of primary cosmic
 rays and hadronic cascading interaction in the atmosphere (~\cite{icrc0171},~\cite{icrc0220},~\cite{icrc0429}).
By muons alone, one can study the cosmic ray anisotropy, or search for point sources, or study the mass composition
(~\cite{icrc1131},~\cite{icrc1140},~\cite{icrc1339}). What concerns model studies,
 the muon charge ratio is of particularly interesting, as it is closely
 related to the hadronic production and decay of the charged pions and kaons in the forward region, and 
 it is sensitive to the primary composition. The energy dependent muon charge ratio from a few hundred
 MeV to muti-TeV has been well measured and shows generally good agreement with the current model 
prediction within experimental errors for muon energies below 1TeV
(~\cite{icrc0128},~\cite{icrc0141},~\cite{icrc0191},~\cite{icrc0921},~\cite{icrc1057}).
When the muon energy is above 1TeV,
 observation shows that model prediction underestimates the muon charge ratio ~\cite{icrc0177}.
With the muon charge ratio at higher energy, MC simulation shows that WILLI-EAS will have a rather
good sensitivity in testing the hadronic interaction models up to PeV energy ~\cite{icrc0128}.

Atmospheric muons originate in the decay of charged pions and kaons produced in the hadronic cascading
 interaction. The intensity of the muon flux is determined by two competing processes, the decay of the
 mesons and the interaction of mesons with the atmospheric nuclei. The lower the effective atmospheric
 temperature, the higher the atmospheric density, the more chance for interaction to happen compared to
 decay, and less is the observed muon flux. This sensitive correlation between atmospheric
 temperature and muon flux makes it a useful tool in monitor the atmospheric variation
(~\cite{icrc0191},~\cite{icrc0192})
 and probe
 the Antarctic ozone hole dynamics and temporal behaviour of the stratospheric temperature
(~\cite{icrc1060},~\cite{icrc1398}).

\begin{figure*}[!t]
  \centering
  \includegraphics[width=14cm,angle=0]{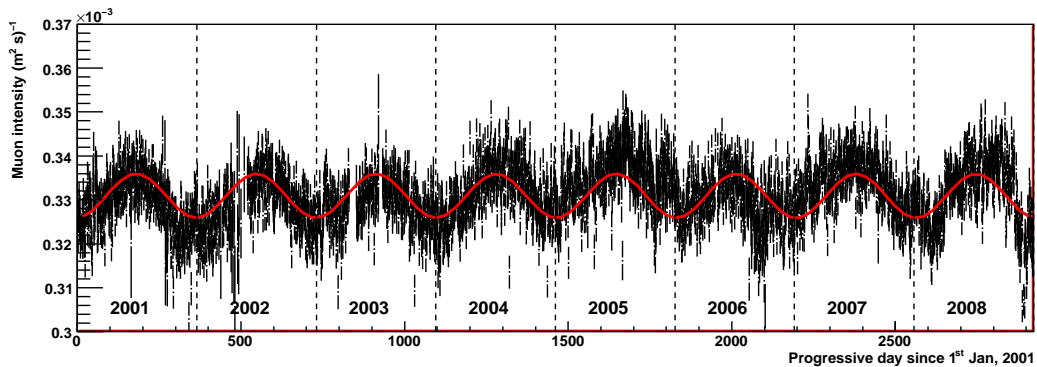}
  \caption{\label{fig8}Annual muon flux modulation for energies great than 1.3 TeV observed by the LVD experiment.
           The amplitude is $1.5 \pm 0.1$ \% and with a phase of $185 \pm 15$ days, with maximum in early July.
           Details can be found in ~\cite{icrc0766}.}
 \end{figure*}

\begin{figure*}[!t]
  \centering
  \includegraphics[width=14cm,angle=0]{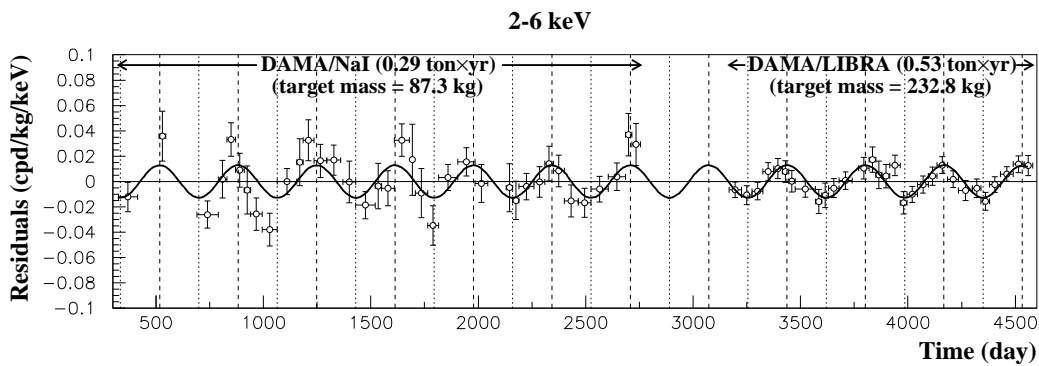}
  \caption{\label{fig9}Annual modulation observed by DAMA NaI and DAMA/LIBRA  experiments 
            with recoil energy between 2-6keV. The amplitude
           is $(0.0129 \pm 0.0016)$ cpd/kg/keV against an overall 
           background counting rate of about 1 cpd/kg/keV
          (corresponding to a relative amplitude of about $ 1.3 \pm 0.1$ \%).
           The phase is $144 \pm 8$ days, with maximum in early June. Details can be found in ~\cite{dama2008}.}
 \end{figure*}

 Due to the very strong penetration power, atmospheric muons are the most important background for
 underground experiments, such as the direct DM search and neutrino experiments. The LVD experiment
 presented a very clear annual muon flux modulation based on 8 years of data for muon energies
 great than 1.3TeV.
 As shown in Fig.\ref{fig8}, the amplitude of the modulation is about $(1.5 \pm 0.1)$ \% and the maximum intensity
 happens in early July ~\cite{icrc0766}. Probably a pure coincident, both amplitude and phase are very close
 to what has been observed by DAMA/NaI and DAMA/LIBRA
 experiment in the same underground laboratory ( Fig.\ref{fig9}, and ~\cite{dama2008} for details).
 It would be of interest for both LVD and DAMA to perform an analysis of the daily modulation. The
 daily modulation should be of similar significance as the annual modulation of
 atmospheric muon but not so the modulation due to the DM interaction.
 As the temperature difference between winter and
 summer is only about twice that of day and night, while in case of DM induced modulation,
 the relevant velocity for annual modulation
 is about 30 km/s in comparison to the day and night effect of this velocity due to the rotation
 of the earth, which is only 1 km/s, a small effect.

 The ANTARES experiment presented muon flux as a function of water depth from 2000
 to 7000 m, this preliminary result agrees with previous measurements by other experiments.
OPERA also reported their preliminary results on the underground muon spectrum at a conference
talk ~\cite{opera_slide}.

Parameterization work on muon flux and spectrum in deep water or ice for the purpose of fast simulation
 have been presented  (~\cite{icrc0642},~\cite{icrc0738}).
 Predictions on muon flux and muon charge ratio up to PeV energy have also been calculated ~\cite{icrc0962}.

Given the fact that the muon component can be used to effectively discriminate the gamma ray showers from
 the overwhelming cosmic ray background showers, the Tibet AS$\gamma$ experiment ~\cite{icrc0297}
 and GRAPSE-3 ~\cite{icrc1180} both decided to
 build or enlarge their muon detector in the near future. With which, the sensitivity of 
gamma ray observation will be improved
 by a factor of ten for Tibet AS$\gamma$ in 100TeV energy range and by a factor of two for GRAPSE-3 in
 multi-TeV energy. In addition, a conceptional proposal ~\cite{icrc0436}
on a $km^{2}$ complex array at high altitude is
 presented for gamma ray astronomy observation with very interesting sensitivity. The center dense
 array and surrounding sparse array will be responsible for low and high energy, respectively. Each
 of the $100 m^{2}$ unit detector towers will contain one electromagnetic layer at the top and two muon layers
 at the bottom, and the measured muon number will be used to reject the cosmic ray background and thus
 improve the sensitivity for the $\gamma$ ray observation.

\section{Summary }
 More progress has been made in the study of the cosmic ray spectrum and more hints for a single source
 model are presented. The compelling evidence for the origin of the galactic cosmic rays is not yet
 found and this situation imposes a strong demand and great interest in precisely measuring
 the mass composition of cosmic rays around the knee energy which heavily relies on the progress in
 understanding the hadronic interaction models. With the ongoing and upcoming activities, we 
 are quite confident that many of the problems will be resolved soon.

\section{Acknowledgments}
I would like to thank the organizers of the ICRC2009 for inviting me to give the
Rapporteur talk and for their hospitality during my stay.
I would like to thank L.K.Ding, Y.Q.Ma, M.Giller, M.Shibata and M.Takita for their very helpful comments,
and thank B.Wang, J.L.Zhang, Q.Yuan, Y.Li, Y.Q.Guo, Z.Y.Feng,I.De Mitri and M.Selvi
 for making calculations, plots or comments. My full gratitude also goes to T. Kozanecki
who helped a lot in smoothing my english writing.
I also wish to thank Wei Wang for his long term help with my research work.
 The work is supported by Natural Sciences Foundation of China (Nos. 10725524 and 10675134),
 and by the Chinese Academy of Sciences (Nos.KJCX2-YW-N13)

\end{document}